\documentclass[10pt,a4paper]{article}

\usepackage{amsmath,amssymb}
\usepackage[pdftex]{graphicx}
\usepackage{latexsym}
\usepackage{slashed}

\makeatletter
\@addtoreset{equation}{section}
\makeatother

\usepackage[top=30truemm,bottom=30truemm,left=25truemm,right=25truemm]{geometry}

\graphicspath{{./fig/}}
\title{Inflation from the Finite Scale Gauged Nambu-Jona-Lasinio Model}
\author{Tomohiro Inagaki${}^{1,2}$, Sergei D. Odintsov${}^{3,4}$ and Hiroki Sakamoto${}^5$\\ \\
\normalsize ${}^1$Information Media Center, Hiroshima University, Higashi-Hiroshima, 739-8521, Japan,\\
\normalsize ${}^2$Core of Research for the Energetic Universe, Hiroshima University, Higashi-Hiroshima, \\
\normalsize 739-8526, Japan,\\
\normalsize ${}^3$ICREA, Passeig Luis Companys,
23, 08010 Barcelona, Spain,\\
\normalsize ${}^4$Institute of Space Sciences (IEEC-CSIC),
Cr.Can Magrans s/n, 08193 Barcelona, Spain,\\
\normalsize ${}^5$Department of Physics, Hiroshima University, Higashi-Hiroshima, 739-8526, Japan}
\begin{document}
\maketitle
%----------------------------------------------------------------------------%

%----------------------------------------------------------------------------%
\begin{abstract}
The possibility to construct an inflationary universe scenario for the finite-scale gauged Nambu-Jona-Lasinio model is investigated. This model can be described by the Higgs-Yukawa type interaction model with the corresponding compositeness scale. Therefore, the one-loop Higgs-Yukawa effective potential is used with the compositeness condition for the study of inflationary dynamics. We evaluate the fluctuations in the cosmic microwave background for the model with a finite compositeness scale in the slow-roll approximation. We find the remarkable dependence on the gauge group and the number of fermion flavors. It is also proved that the model has similar behavior with the $\phi^{4n}$ chaotic inflation and the Starobinsky model at the flat and steep limits, respectively. It is demonstrated that realistic inflation consistent with Planck data is possible for a range of theory parameters.
%%%We found these behavior can be distinguished under the consideration of infinity limit of the compositeness scale.
 \end{abstract}
%----------------------------------------------------------------------------%

%----------------------------------------------------------------------------%
\section{Introduction}
The occurrence of the early-time inflation provides eventually the easiest explanation of the astrophysical data, including that the large-scale universe is approximately isotropic, homogeneous and spatially-flat.
The possible origin of the inflationary era maybe related with existing particle physics models (for review, see \cite{Linde:2007fr, Martin:2013tda, Linde:2014nna, Gorbunov:2011zzc}).
The quantum fluctuations in the particle physics theory grow up through the inflationary expansion. It can be considered as the origin of the fluctuations in the cosmic microwave background (CMB). Therefore, the cosmological predictions of particle physics theory can be inspected by the Planck observational data.

 A composite scalar field maybe one of the interesting candidates to provide the potential energy for the inflationary expansion. The Nambu-Jona-Lasinio (NJL) model \cite{Nambu:1961} describes  the composite scalar field which appears in the low-energy phenomena. The gauged NJL model has been introduced as a low energy-effective model of QCD with QED gauge interaction. The scale up model can be used as a prototype model of a composite scalar field theory with a gauge interaction at high energy. Other composite models have been also studied as the origin to induce the inflationary expansion (see, for example \cite{Channuie:2011rq, Bezrukov:2011mv, Channuie:2012bv, G.:2015rgn}).

In Ref.~\cite{Inagaki:2015eza} the CMB fluctuations are evaluated in the gauged NJL model. The gauged NJL model can be represented as the gauged Higgs-Yukawa theory below the compositeness scale (see, for example \cite{Miransky:1993, Harada:2003jx, Kondo:1991yk, Kondo:1992sq, Kondo:1993jq, Leung:1985sn, Leung:1989hw, Harada:1994wy}). In the previous study the CMB fluctuations are calculated in the gauged Higgs-Yukawa theory with the compositeness scale.
The consistency with the Planck 2015 data has been shown at the infinite compositeness scale limit.

In the standard scenario of the chaotic inflation it is often assumed that the field  starts from a super-Planckian domain. Although the infinite compositeness scale limit \cite{Inagaki:2015eza} reduces the model parameters, the initial field value is too large to adopt the infinite compositeness scale limit. A more realistic situation corresponds to the model with a finite compositeness scale. Therefore, it is quite natural to investigate the inflation induced by the gauged NJL model with a finite compositeness scale and calculate the CMB fluctuations in such a case.
%%%
Therefore we find alternative features of the gauged NJL model.

This paper is organized as follows. Section~2 is devoted to the review of the gauged NJL model. Using the auxiliary field method and applying the renormalization group improvement technique, the model is rewritten as the gauged Higgs-Yukawa theory. In Sec.~3, we employ the slow-roll approximation and present the explicit forms of the parameters for CMB fluctuations. In Sec.~4, we find the analytic solutions of the CMB fluctuations at the flat and steep limits. It is observed that the gauged NJL model behaves as the $\phi^{4n}$ chaotic inflation and Starobinsky model, respectively at the flat and steep limits.
%%%
We also discuss why such a feature can not be observed at the infinite compositeness scale limit considered in Ref.~\cite{Inagaki:2015eza}. In Sec.~5 we numerically calculate the CMB fluctuations and confirm their consistency with the analytic results. The correspondence with Planck data is explicitly established. The reheating temperature is roughly estimated in Sec.~6. Finally, some concluding remarks are given.
%----------------------------------------------------------------------------%

%----------------------------------------------------------------------------%
\section{Gauged NJL model with a finite compositeness scale}

Following the procedure developed in Refs.~\cite{Inagaki:2015eza, Geyer:1996wg, Geyer:1996kg}, let us construct the effective potential for the composite scalar field at inflationary era. The gauged NJL model is employed as a prototype model to generate the composite scalar field.
Thus, we start from the Lagrangian of the gauged NJL model, i.e. an $SU(N_c)$ gauge theory with four-fermion interactions,
\begin{align}
        {\cal L}_{gNJL} = {\cal L}_{gauge} + \bar{\psi} i\hat{\slashed{D}}\psi + \frac{16\pi^2 g_4}{8 N_f N_c \Lambda^2} \left[\left(\bar{\psi}\psi\right)^2+\left(\bar{\psi}i\gamma_5 \tau^{a}\psi\right)^2\right] ,
        \label{L:gNJL}
\end{align}
where ${\cal L}_{gauge}$ shows the Lagrangian of the pure $SU(N_c)$ gauge theory, $\hat{D}$ is the covariant derivative and $\tau^a$ are the generators of the $SU(N_f)$ flavor symmetry. The four-fermion coupling is composed by the dimensionless parameter $g_4$ and compositeness scale $\Lambda$. According to the auxiliary field method, the Lagrangian \eqref{L:gNJL} is rewritten as
\begin{align}
        {\cal L}_{aux} = {\cal L}_{gauge} + \bar{\psi} \left(i\hat{\slashed{D}}-\sigma-i\gamma_5 \tau^a\pi^a \right)\psi - \frac{2 N_f N_c\Lambda^2}{16\pi^2 g_4} \left(\sigma^2+{\pi^a}^2\right),
        \label{L:auxi}
\end{align}
where $\sigma$ and $\pi^a$ represent the composite scalar and pseudo-scalar fields.

From the other side, the corresponding (renormalizable) gauge-Higgs-Yukawa Lagrangian is given by
\begin{align}
        {\cal L}_{gHY} = &{\cal L}_{gauge} + \frac{1}{2y^2}\partial_\mu\sigma\partial^\mu\sigma
        + \frac{1}{2y^2}\partial_\mu\pi^a\partial^\mu\pi^a
        - \frac{1}{2}\frac{m^2}{y^2}(\sigma^2+\pi^a\pi^a) - \frac{\lambda}{4y^4}(\sigma^2+\pi^a\pi^a)^2 \nonumber \\
        &-\frac{1}{2}\frac{\xi}{y^2} R(\sigma^2+\pi^a\pi^a) + \bar{\psi} i\hat{\slashed{D}}\psi -  \bar{\psi}(\sigma +i\gamma_5 \tau^a\pi^a )\psi,
        \label{L:gHY}
\end{align}
where $y$ indicates the Yukawa coupling. Here we rescale the fields in the ordinary gauge-Higgs-Yukawa model to $\sigma \to \sigma/y$ and $\pi^a \to \pi^a/y$.
There is a solution of the renormalization group (RG) equations where the Lagrangian \eqref{L:auxi} has an equivalent form with \eqref{L:gHY} at the compositeness scale $\Lambda$. The compositeness of the fields, $\sigma$ and $\pi^a$, can be represented by the conditions \cite{Bardeen:1989ds, Hill:1991jc},
\begin{align}
        \frac{1}{y^2(t_\Lambda)} = 0,\quad
        \frac{\lambda(t_\Lambda)}{y^4(t_\Lambda)} = 0,\quad
        \frac{m^2(t_\Lambda)}{y^2(t_\Lambda)} = \frac{2a}{16\pi^2}\Lambda^2\left(\frac{1}{g_4}-\frac{1}{\Omega(t_\Lambda)}\right),\quad
        \xi(t_\Lambda) = \frac{1}{6},
        \label{cond:Composite}
\end{align}
where $a = 2N_cN_f$ and $\Omega(t_\Lambda)$ is a function of $t \equiv \ln(\mu/\mu_0)$ with the renormalization scale, $\mu$, and the reference scale, $\mu_0$, at $\mu = \Lambda$. Below we suppose only the composite scalar $\sigma$ contributes to the evolution of the Universe and drop the pseudo-scalar field $\pi^a$.

The running Yukawa coupling $y(t)$, the quartic scalar coupling $\lambda(t)$ and the curvature coupling $\xi(t)$ in the low energy effective model of the gauged NJL theory are found by solving the RG equations with the boundary conditions (\ref{cond:Composite}). Here we neglect running of the $SU(N_c)$ gauge coupling $g$, for simplicity. Thus, the function $\Omega(t_\Lambda)$ coincides with $w = 1 - \alpha/(2\alpha_c)$, where $\alpha \equiv g^2/(4\pi)$ and $\alpha_c \equiv 2\pi N_c/(3N_c^2-3)$. The solutions of RG equations are given by
\begin{align}
        &y^2(t) = \frac{16\pi^2}{2a}\frac{\alpha}{\alpha_c}\left[1 - \left(\frac{\mu^2}{\Lambda^2}\right)^{1-w}\right]^{-1},
\label{running:y:b0}\\
        &\frac{\lambda(t)}{y^4(t)} = \frac{2a}{16\pi^2}\frac{\alpha_c}{\alpha}\left[1 - \left(\frac{\mu^2}{\Lambda^2}\right)^{2-2w}\right],
\label{running:lambda:b0}\\
        &\xi(t) = \frac{1}{6}.
        \label{running:xi}
\end{align}

We introduce the compositeness conditions into the one-loop effective potential of the gauge-Higgs-Yukawa model up to linear in the curvature terms and obtain
\begin{align}
        V^{1 loop}(\sigma) = \frac{1}{2}m^2\sigma^2 + \frac{1}{4}\lambda\sigma^4 + \frac{1}{2}\xi R\sigma^2 - \frac{ay^4\sigma^4}{2\cdot 16\pi^2}\left(\ln\frac{y^2\sigma^2}{\mu^2} - \frac{3}{2}\right) - \frac{aRy^2\sigma^2}{12\cdot 16\pi^2}\left(\ln\frac{y^2\sigma^2}{\mu^2} - 1\right).
\label{effpot:oneloop}
\end{align}
The solution of the RG equation for effective potential satisfies
\begin{align}
        V(g,y,\lambda,m^2,\xi,\sigma,\mu) = V(\bar{g}(t),\bar{y}(t),\bar{\lambda}(t),\bar{m}^2(t),\bar{\xi}(t),\bar{\sigma}(t),\mu e^t),
\label{RG:effpot}
\end{align}
where the barred quantities represent the renormalized ones at the scale $\mu e^t$. This scale is fixed to drop the logarithmic term in RG invariant effective potential,
\begin{align}
        e^t=\left(\frac{y\sigma}{\mu}\right)^{1/(2-w)}.
        \label{scale:mu}
\end{align}

From the one-loop effective potential \eqref{effpot:oneloop} with compositeness conditions \eqref{cond:Composite} and \eqref{scale:mu}, one can evaluate the RG improved effective potential \cite{Harada:1994wy,Geyer:1996wg},
\begin{align}
        V = \frac{B}{2}\sigma^2 + \frac{C_1}{4}\sigma^{4/(2-w)} - \frac{C_2}{4}\sigma^4 + \frac{R}{2}\frac{D_1}{6}\sigma^{2/(2-w)} - \frac{R}{2}\frac{D_2}{6}\sigma^2,
\end{align}
where the functions $B$, $C_1$, $C_2$, $D_1$ and $D_2$ are given by
\begin{align}
        &B = 2(1-w)\Lambda^2\left(\frac{1}{g_{4}(\Lambda)} - \frac{1}{w}\right)\frac{(\mu^2/\Lambda^2)^{1-w}}{1-(\mu^2/\Lambda^2)^{1-w}}, \\
        &C_1 = \frac{a\mu^4}{16\pi^2}\left(\frac{4-3w}{1-w}\right)\left[\frac{16\pi^2}{a\mu^2}\frac{1-w}{1-(\mu^2/\Lambda^2)^{1-w}}\right]^{2/(2-w)}, \\
        &C_2 = \frac{16\pi^2}{a}(1-w)\left[\frac{(\mu^2/\Lambda^2)^{1-w}}{1-(\mu^2/\Lambda^2)^{1-w}}\right]^{2}, \\
        &D_1 = \frac{a\mu^2}{16\pi^2}\left(\frac{2-w}{1-w}\right)\left[\frac{16\pi^2}{a\mu^2}\frac{1-w}{1-(\mu^2/\Lambda^2)^{1-w}}\right]^{1/(2-w)}, \\
        &D_2 = \frac{(\mu^2/\Lambda^2)^{1-w}}{1-(\mu^2/\Lambda^2)^{1-w}}.
\end{align}
It should be noted that the functions, $C_2$ and $D_2$, disappear at the large $\Lambda$ limit \cite{Inagaki:2015eza}. Note that RG improved effective potential of renormalizable gauge theory in curved spacetime is found long ago \cite{Elizalde:1993ee, Elizalde:1993ew}. Such RG improved potential has been used to study the (elementary scalar) inflation in refs.\cite{DeSimone:2008ei, Mukaigawa:1997nh, Lee:2013nv, Okada:2013vxa, Barenboim:2013wra, Woodard:2008yt, Ren:2014sya, Hamada:2014xka, Inagaki:2014wva, Inagaki:2015fva, Elizalde:2014xva, Hamada:2014wna, He:2014ora, Herranen:2015aja,Takahashi:2016uyq, Herranen:2016xsy, Kannike:2015kda, Artymowski:2016dlz}.

Therefore, we obtain the effective action for the composite scalar with Einstein-Hilbert term,
\begin{align}
        S = \int d^4x \sqrt{-g} \left[-\frac{1}{2}R + \frac{1}{2}g^{\mu\nu}\partial_\mu\sigma\partial_\nu\sigma - V\right],
        \label{Jordan frame}
\end{align}
where  the reduced Planck unit $M_{p}^2 = 1/(8\pi G) = 1$ is used.

%----------------------------------------------------------------------------%
\section{CMB fluctuations in the gauged NJL inflation}
Let us evaluate the CMB fluctuations starting from the effective action \eqref{Jordan frame}.
Below the compositeness scale the composite scalar field couples with the curvature $R$.
The useful way to evaluate the CMB fluctuations is to transform the non-minimal (Jordan frame) description to the minimal  (Einstein frame) description.
\footnote {It is known that in slow-roll approximation the calculation of primordial spectral indices gives effectively the same numerical prediction for these indices calculated in Jordan or in Einstein frames \cite{Kaiser:1994vs, Kaiser:2015usz, Domenech:2015qoa, Brooker:2016oqa, Kamenshchik:2014waa}.}

 This is achieved by the Weyl transformation for the metric tensor $g_{\mu\nu}$
\begin{align}
\tilde{g}_{\mu\nu} = \Omega^2(x) g_{\mu\nu},
\end{align}
where the Weyl factor $\Omega^2(x)$ is tuned to cancel out the non-minimal curvature coupling term.

After the Weyl transformation and the redefinition of the field one obtains the action with a canonical kinetic term in the Einstein frame,
\begin{align}
        S_E = \int dx^4 \sqrt{-\tilde{g}} \left[-\frac{1}{2}\tilde{R} + \frac{1}{2}\tilde{g}^{\mu\nu}\partial_\mu \phi \partial_\nu \phi - V_E\right],
\end{align}
where the subscript $E$ indicates the quantities in the Einstein frame, and the redefined scalar field $\phi$ satisfies
\begin{align}
        \frac{d\phi}{d\sigma} = \sqrt{\frac{f}{\Omega^2}}, \quad f \equiv 1 + \frac{3}{2}\frac{(\Omega^2_\sigma)^2}{\Omega^2}.
        \label{eq:redef}
\end{align}
Here and hereafter we use the subscripts $\sigma$ and $\phi$ to describe the derivative with respect to them.

In the Einstein frame the Weyl factor is found to be
\begin{align}
        \Omega^2 = 1 + \zeta_1 \mu^{2-2n} \sigma^{2n} - \zeta_2\sigma^2,
        \label{weyl factor}
\end{align}
 and the effective potential is given by
\begin{align}
        V_E = \frac{U}{\Omega^4}, \quad U = \lambda_1 \mu^2 \sigma^2 + \lambda_2 \mu^{4(1-n)} \sigma^{4n} - \lambda_3 \sigma^4,
        \label{einstein frame}
\end{align}
where we set
\begin{align}
        & \zeta_1 = \frac{1}{6n}\left(\frac{a}{16\pi^2}\frac{n}{1-n}\right)^{1-n}\left[\frac{1}{1 - (\mu^2/\Lambda^2)^{\frac{1-n}{n}}}\right]^n, \\
        &\zeta_2 = \frac{1}{6}\left[\frac{({\mu}^2/{\Lambda}^2)^{\frac{1-n}{n}}}{1-({\mu}^2/{\Lambda}^2)^{\frac{1-n}{n}}}\right], \label{eq:zeta2}\\
        &\lambda_1 = \frac{1-n}{nG_{4r}}\left[\frac{1}{1 - (\mu^2/\Lambda^2)^{\frac{1-n}{n}}}\right], \\
        &\lambda_2 = \frac{3-2n}{4n}\left(\frac{a}{16\pi^2}\frac{n}{1-n}\right)^{1-2n}\left[\frac{1}{1 - (\mu^2/\Lambda^2)^{\frac{1-n}{n}}}\right]^{2n}, \\
        &\lambda_3 = \frac{4\pi^2}{a}\frac{1-n}{n} \left[\frac{(\mu^2/\Lambda^2)^{\frac{1-n}{n}}}{1-(\mu^2/\Lambda^2)^{\frac{1-n}{n}}}\right]^2, \label{eq:lambda3}
        \end{align}
and $1/G_{4r}$ shows the renormalized coupling, $1/G_{4} \equiv 1/g_{4}(\Lambda) - 1/w$ with $(\mu^2/\Lambda^2)^{-w}$, and $n = 1/(2-w)$.

The stability of the potential is evaluated by observing the positivity of the first derivative, $\displaystyle \frac{\partial V}{\partial \sigma}>0$. Below we evaluate the CMB fluctuations starting from $\sigma$ where the potential satisfies the stability condition.
For $\mu/\Lambda \ll 1$ the stability condition is given by
\begin{align}
  a<\frac{16\pi^2(1-n)}{n}\frac{(\sigma^2)^{\frac{n-1}{n}}}{\mu^2}\left[ \frac{(9-6n)n^2G_{4r}}{(2n-1)(1-n)}\right]^{1/n}.
\end{align}
At the limit, $\alpha \rightarrow 0$ ($n \rightarrow 1$), it reduces to a simple form
\begin{align}
  a<\frac{48\pi^2}{\mu^2}G_{4r}.
\end{align}
It reproduces the one obtained in Ref.~\cite{Inagaki:2015eza}, $N_f N_c<24\pi^2$ for $\mu=1$ and $G_{4r}=1$.

In the analysis of the CMB fluctuations the parameter $n$ in the exponent of Eqs.~(\ref{weyl factor}) and (\ref{einstein frame}) has a decisive role. It is given by
\begin{align}
        n=\frac{4\pi N_c}{4\pi N_c+3\alpha(N_c^2-1)} \xrightarrow{N_c^2\gg 1} \frac{4\pi }{4\pi +3\alpha N_c} .
        \label{def:n}
\end{align}
At the limit $N_c^2\gg 1$ the parameter $n$ is described as a function of $\alpha N_c$. On the other hand the RG equations
(\ref{running:y:b0}), (\ref{running:lambda:b0}) and (\ref{running:xi}) are calculated in the perturbative regime $0<\alpha N_c < 1$ with the large $N_c$ limit. It means that a non-perturbative approach is necessary for
\begin{align}
        n \lesssim \frac{1}{1+3/(4\pi)} \sim 0.8 .
\end{align}

%----------------------------------------------------------------------------%
%\section{Slow-roll parameters and CMB fluctuations}
The fluctuations of cosmological microwave background (CMB)  led to the quantum fluctuations of the inflaton. Thus, the CMB fluctuations are considered as eventual evidence of the inflationary expansion.
 In the Einstein frame the dynamics of the inflaton $\phi$ is described by
\begin{align}
        \ddot{\phi} + 3H\dot{\phi} = -\frac{\partial V_E}{\partial \phi}, \\
        H^2 = \frac{1}{3}\left(\frac{1}{2}\dot{\phi}^2 + V_E\right),
\end{align}
where $H$ is the Hubble parameter and we use the dot for the time derivative.

We assume that the inflaton rolls down slowly from the initial value on its potential,
\begin{align}
        \dot{\phi} \ll V_E, \quad \ddot{\phi} \ll \left|\frac{\partial V_E}{\partial \phi}\right|.
\end{align}
and employ the slow-roll approximation.
It is more convenient for practical calculations to define the slow-roll parameters, $\varepsilon (\ll 1)$, $\eta (\ll 1)$ and $\xi_{CMB}$,
\begin{align}
        & \varepsilon \equiv \frac{1}{2}\left(\frac{1}{V_E}\frac{\partial V_E}{\partial \phi}\right)^2, \\
        & \eta \equiv \frac{1}{V_E}\frac{\partial^2 V_E}{\partial \phi^2},
        \label{eq:slow-roll parameter} \\
        & \xi_{CMB} \equiv \frac{1}{V_E^2}\frac{\partial V_E}{\partial \phi}\frac{\partial^3 V_E}{\partial \phi^3}.
        \label{eq:xi_cmb}
\end{align}
Substituting the effective potential (\ref{einstein frame}) with the Weyl factor (\ref{weyl factor}), we obtain
\begin{align}
        \varepsilon &= \frac{1}{2}\left(\frac{V_{E\phi}}{V_E}\right)^2 = \frac{1}{2}\frac{\Omega^2}{f}\left(\frac{U_\sigma}{U} - 2\frac{\Omega^2_\sigma}{\Omega^2}\right)^2, \label{epsilon:U}
\end{align}
\begin{align}
        \eta &= \frac{V_{E\phi\phi}}{V_E} \nonumber \\
        &= \frac{\Omega^2_\sigma f - \Omega^2 f_\sigma}{2f^2}\left(\frac{U_\sigma}{U} - 2\frac{\Omega^2_\sigma}{\Omega^2}\right) + \frac{\Omega^2}{f}\left[\frac{U_{\sigma\sigma}}{U} - \left(\frac{U_\sigma}{U}\right)^2 - 2\frac{\Omega^2_{\sigma\sigma}}{\Omega^2} + 2\left(\frac{\Omega^2_\sigma}{\Omega^2}\right)^2\right] + \frac{\Omega^2}{f}\left(\frac{U_\sigma}{U} - 2\frac{\Omega^2_\sigma}{\Omega^2}\right)^2, \label{eta:U}
\end{align}
\begin{align}
        \xi_{CMB} &= \frac{V_{E\phi}V_{E\phi\phi\phi}}{V^2_E} \nonumber \\
        &= \frac{\Omega^8}{f^4}\left[\frac{1}{2}\left(\frac{U_\sigma}{U} - 2\frac{\Omega^2_\sigma}{\Omega^2}\right)^2 \left(\frac{\Omega^2_{\sigma\sigma}}{\Omega^2} - \frac{2\Omega^2_\sigma f_\sigma}{\Omega^2 f} - \frac{f_{\sigma\sigma}}{f} + 2\frac{f^2_\sigma}{f^2}\right)\right. \nonumber \\
        &+ 2\left(\frac{U_\sigma}{U} - 2\frac{\Omega^2_\sigma}{\Omega^2}\right)\left(\frac{U_{\sigma\sigma}}{U} - 4\frac{U_\sigma}{U}\frac{\Omega^2_\sigma}{\Omega^2} - 2\frac{\Omega^2_{\sigma\sigma}}{\Omega^2} + 6\frac{\Omega^4_\sigma}{\Omega^4}\right) \left(\frac{\Omega^2_\sigma}{\Omega^2} - \frac{f_\sigma}{f}\right) \nonumber \\
        &- \frac{1}{2}\left(\frac{U_\sigma}{U} - 2\frac{\Omega^2_\sigma}{\Omega^2}\right)^3 \left(\frac{\Omega^2_\sigma}{\Omega^2} - \frac{f_\sigma}{f}\right) \left(2-\frac{f^2}{\Omega^4}\right) \nonumber \\
        &+ \left(\frac{U_{\sigma\sigma}}{U} - 4\frac{U_\sigma}{U}\frac{\Omega^2_\sigma}{\Omega^2} - 2\frac{\Omega^2_{\sigma\sigma}}{\Omega^2} + 6\frac{\Omega^4_\sigma}{\Omega^4}\right)^2 \nonumber \\
        &- \left(\frac{U_\sigma}{U} - 2\frac{\Omega^2_\sigma}{\Omega^2}\right)^2 \left(\frac{U_{\sigma\sigma}}{U} - 4\frac{U_\sigma}{U}\frac{\Omega^2_\sigma}{\Omega^2} - 2\frac{\Omega^2_{\sigma\sigma}}{\Omega^2} + 6\frac{\Omega^4_\sigma}{\Omega^4}\right) \left(2-\frac{f^2}{\Omega^4}\right) \nonumber \\
        &+ \left.\left(\frac{U_\sigma}{U} - 2\frac{\Omega^2_\sigma}{\Omega^2}\right) \left(\frac{U_{\sigma\sigma\sigma}}{U} - 6\frac{U_{\sigma\sigma}}{U}\frac{\Omega^2_\sigma}{\Omega^2} - 6\frac{U_\sigma}{U}\frac{\Omega^2_{\sigma\sigma}}{\Omega^2} + 18\frac{U_\sigma}{U}\frac{\Omega^4_\sigma}{\Omega^4} - 2\frac{\Omega^2_{\sigma\sigma\sigma}}{\Omega^2} + 18\frac{\Omega^2_\sigma\Omega^2_{\sigma\sigma}}{\Omega^4} - 24\frac{\Omega^6_\sigma}{\Omega^6}\right)\right],\label{xi:U}
\end{align}
where the subscript $\sigma\sigma$ and $\sigma\sigma\sigma$ indicates the second and third derivative with respect to $\sigma$.

We assume that the volume of the Universe increases $e^N$ times during the inflation. Thus, the e-folding number $N$ is given by
\begin{align}
        N = \int^{\phi_N}_{\phi_{end}}\frac{V_E}{\partial V_E/\partial \phi}d\phi,
        \label{eq:e-folds}
\end{align}
where $\phi_N$, $\phi_{end}$ denote the field configuration of the horizon crossing and the end of the inflation, respectively.
Inserting the effective potential (\ref{einstein frame}), the e-folding number is written as
\begin{align}
        N = \int^{\sigma_N}_{\sigma_{end}} \frac{f}{\Omega^2}\frac{1}{U_\sigma/U - 2\Omega^2_\sigma/\Omega^2}d\sigma.
\label{N:U}
\end{align}

Planck satellite collaboration has observed the curvature perturbation $A_s$, the spectral index $n_s$, the running of the spectral index $\alpha_s$ and the tensor-to-scalar ratio $r$. In the slow-roll approximation these observables are given by \cite{Kohri:2013mxa}
\begin{align}
        & A_s = \left.\frac{V_E}{24\pi^2\varepsilon}\right|_{\sigma=\sigma_N}, \label{eq:As}\\
        & n_s = 1 + 2\eta - \left. 6\varepsilon\right|_{\sigma=\sigma_N}, \label{eq:ns}\\
        & \alpha_s = \left.\frac{dn_s}{d\ln k}\right|_{\sigma=\sigma_N} = -24\varepsilon^2 + 16\varepsilon\eta - \left. 2\xi_{CMB}\right|_{\sigma=\sigma_N}, \label{eq:alphas}\\
        & r = \left. 16\varepsilon\right|_{\sigma=\sigma_N}. \label{eq:r}
\end{align}
The observational values of these parameters are given as $\ln(10^{10} A_s) = 3.094\pm0.034$, $n_s = 0.9645\pm0.0049$, $\alpha_s = -0.0057\pm0.0071$ (68\% CL, Planck TT,TE,EE+lowP) and $r_{0.002} < 0.10$ (95\% CL, Planck TT,TE,EE+lowP) \cite{Ade:2015lrj} .
%----------------------------------------------------------------------------%

\section{CMB fluctuations at the flat and steep limits}

CMB fluctuations for gauged NJL model are evaluated by substituting the slow-roll parameters at the horizon crossing to \eqref{eq:As}, \eqref{eq:ns}, \eqref{eq:alphas} and \eqref{eq:r}.
In order to analytically evaluate the CMB fluctuations we assume, only this section, $\mu/\Lambda \ll 1$ and ignore the last term in \eqref{weyl factor} and \eqref{einstein frame}. Here, we consider the following two limits: $\Omega^2 \gg (\Omega^2_\sigma)^2$ and $\Omega^2 \ll (\Omega^2_\sigma)^2$. The former case is named the flat limit and the latter case is called the steep limit \cite{Mosk:2014cba}. These limits characterize the specific behavior of the inflation.

\subsection{Flat limit ($\Omega^2\gg (\Omega_\sigma^2)^2$)}
Let us   assume that the renormalization scale $\mu$ is less than Planck scale, i.e. $\mu<1$ in our notations.
For $2n < 1$ the first term of $U$ in \eqref{einstein frame} dominates the energy density of the inflationary Universe. The first term induces the $\phi^2$ chaotic inflation. Since the second term has a dominant contribution for $2n > 1$, the $\phi^{4n}$ chaotic inflation takes place. At the massless limit, $\lambda_1=0$ the first term disappears and the model is identical to the $\phi^{4n}$ chaotic inflation.

At the flat limit $\Omega^2\gg (\Omega_\sigma^2)^2$, Eqs.~(\ref{epsilon:U}) - (\ref{N:U}) simplify to
\begin{align}
        \varepsilon =& \frac{8n^2}{\sigma^2}, \\
        \eta =& \frac{4n(4n-1)}{\sigma^2} + \frac{16n^2 + 4(1 - 2n)n - 4n(1 + 7n)}{\sigma^2}\zeta_1 \mu^{2-2n} \sigma^{2n}, \\
        \xi_{CMB} =& \frac{16n^2(4n-1)(4n-2)}{\sigma^4} \nonumber \\
        &+ \frac{16n \left[32 n^3-4n^2 (17 n+10)-2n (11n^2-15n-4)-n (4n^2 -10n +1)\right]}{\sigma^4}\zeta_1 \mu^{2-2n} \sigma^{2n},\\
        N=&\frac{\sigma_N^2}{8n},
\end{align}
where all the sub-dominant contributions are ignored.
Substituting these parameters into Eqs.(\ref{eq:As}), (\ref{eq:ns}), (\ref{eq:alphas}) and (\ref{eq:r}), we obtain $A_s$, $n_s$, $r$ and $\alpha_s$ as a function of the model parameters and e-folding number, $N$,
\begin{align}
        A_s &= \begin{cases}
                        \displaystyle \frac{\lambda_1\mu^{2}N^{2}}{3\pi^2} \quad &(2n < 1 \textrm{ and } \lambda_1\neq 0),\\[2mm]
                        \displaystyle \frac{\lambda_2\mu^{4-4n}(8nN)^{2n+1}}{192\pi^2 n^2} \quad &(2n > 1 \textrm{ and/or }\lambda_1= 0),
                \end{cases}
        \\
        n_s &= \begin{cases}
                \displaystyle 1 - \frac{2}{N} \\[2mm]%\quad &(m < 1), \\
                \displaystyle 1 - \frac{2n+1}{N} %\quad &(m > 1 \quad \textrm{or massless limit}),
                \end{cases}
                %\\
                r = \begin{cases}
                        \displaystyle \frac{8}{N} \\[2mm]%\quad &(m < 1), \\
                        \displaystyle \frac{16n}{N} %\quad &(m > 1 \quad \textrm{or massless limit}),
                \end{cases}
                %\\
                \alpha_s = \begin{cases}
                \displaystyle -\frac{2}{N^2} \quad &(2n < 1 \textrm{ and } \lambda_1\neq 0),\\[2mm]
                \displaystyle -\frac{2n+1}{N^2} \quad &(2n > 1 \textrm{ and/or }\lambda_1= 0),
                \end{cases}
                \label{limit:flat}
\end{align}
As is known, the e-folding number $N$ should be in the interval, $50\sim 60$, to solve the flatness problem of the Universe.

To find the typical CMB fluctuations at the flat limit we tune the scale $\mu$ to reproduce the observed value of $A_s$ with $N_cN_f \sim O(1)$ and evaluate $n_s$, $r$ and $\alpha_s$ at $\lambda_1=0$ for $2n = 1.5$, $2n = 1$, and $2n = 0.5$. The results are presented in Table~\ref{tab:tab1}. The result is consistent with the observed values of Planck 2015 for $2n = 0.5$. Such a small $n$ is beyond the scope of the perturbative expansion with respect to $\alpha N_c$. Thus, we conclude that a consistent value for the tensor-to-scalar ratio, $r (<0.10)$, can not be obtained in the perturbative regime at the flat limit .
\begin{table}[htb]
  \centering
  \begin{tabular}{ll|llll}
  &   & $\mu$ & $n_s$ & $r$ & $\alpha_s$\\ \hline
  &  $2n=1.5$ & $O(10^{-13})$ & 0.95 & 0.24 & -0.001\\
   $N$=50  &   $2n=1$ & $O(10^{-7})$ & 0.96 & $0.16$ & -0.0008\\
  &  $2n=0.5$ & $O(10^{-3})$ & 0.97 & 0.08 & -0.0006\\ \hline
    &  $2n=1.5$ & $O(10^{-13})$ & 0.958 & 0.2 & -0.0007 \\
   $N$=60  &   $2n=1$ & $O(10^{-7})$ & 0.97 & 0.13  & -0.0006\\
  &  $2n=0.5$ & $O(10^{-3})$ & 0.975 & 0.07 & -0.0004\\
  \end{tabular}
  \caption{The typical values of the CMB fluctuations for $\lambda_1=0$ and $N_cN_f \sim O(1)$ at the flat limit.}
  \label{tab:tab1}
\end{table}

\subsection{Steep limit ($\Omega^2\ll (\Omega_\sigma^2)^2$)}
Let us now consider the case $1\ll (\Omega_\sigma^2)^2/\Omega^2$. In this case \eqref{eq:redef} reduces to
\begin{align}
        \frac{d\phi}{d\sigma} \sim \sqrt{\frac{3}{2}\left(\frac{\Omega_\sigma^2}{\Omega^2}\right)^2}.
        \label{eq:redef:steep}
\end{align}
Solving this equation with $\Omega^2|_{\phi=0}=1$, one obtains the explicit expression of the Weyl factor,
\begin{align}
        \Omega^2 = \exp\left(\pm\sqrt{\frac{2}{3}}\phi\right).
        \label{Omega:steep}
\end{align}

Here we confine ourselves for the massless limit of the composite scalar field $\lambda_1 =0$. Thus, the potential in the Jordan frame is rewritten as
\begin{align}
  U = \lambda_2 \mu^{4(1-n)} \sigma^{4n} = \hat{\lambda} (\Omega^2 - 1)^2,
  \label{U:steep}
\end{align}
with
\begin{align}
  \hat{\lambda} \equiv \frac{\lambda_2}{\zeta_1^2} = \frac{144\pi^2}{a}(3-2n)(1-n).
\end{align}
Plugging \eqref{Omega:steep} to \eqref{U:steep}, the potential in the Einstein frame is represented as
\begin{align}
  V_E = \frac{U}{\Omega^4} = \hat{\lambda} (1 - \exp(-\sqrt{2/3}\phi))^2.
\end{align}
It is equivalent to the potential in the Starobinsky inflation\cite{sta}. As is known
Starobinsky inflation is consistent with the Planck observational data.

Thus, the slow-roll parameters are given by
\begin{align}
        &\varepsilon = \frac{4}{3}\frac{1}{(e^{\sqrt{2/3}\phi} - 1)^2}, \\
        &\eta = -\frac{4}{3}\frac{(e^{\sqrt{2/3}\phi} - 2)}{(e^{\sqrt{2/3}\phi} - 1)^2}, \\
        &\xi_{CMB} = \frac{16}{9}\frac{(e^{\sqrt{2/3}\phi} - 4)}{(e^{\sqrt{2/3}\phi} - 1)^3},
\end{align}
and the e-folding number is
\begin{align}
        N = \frac{3}{4}\left(e^{\sqrt{2/3}\phi_N} - e^{\sqrt{2/3}\phi_{end}}\right) - \sqrt{\frac{3}{8}} (\phi_N - \phi_{end}).
\end{align}

At the steep limit the spectral index and tensor-to-scalar ratio are approximated as
\begin{align}
        n_s \sim 1 - \frac{2}{N} - \frac{9}{2N^2}, \quad
        r \sim \frac{12}{N^2}, \quad
        \alpha_s \sim - \frac{2}{N^2}.
\label{sol:steep}
\end{align}
These equations reproduce the results of the Starobinsky model.
We present the values, $n_s$, $r$ and $\alpha_s$ for $N=50$ and $60$ in Table~\ref{tab:tab:steep}.
\begin{table}[htb]
  \centering
  \begin{tabular}{l|lll}
  $N$ &   $n_s$ & $r$ & $\alpha_s$\\ \hline
  50  &    0.9582 & $0.0048$ & -0.00080\\
  60  &    0.9654 & 0.0033  & -0.00056\\
  \end{tabular}
  \caption{The CMB fluctuations for $\lambda_1=0$ at the steep limit.}
  \label{tab:tab:steep}
\end{table}

In the present case the curvature perturbation is given by
\begin{align}
  A_s \sim \frac{\hat{\lambda} N^2}{18\pi^2}\left(1 - \frac{3}{4N}\right)^4.
\end{align}
To generate the observed curvature perturbation $A_s$, the effective coupling $\hat{\lambda}$ should be of the order $O(10^{-10})$.

It should be noticed that the solution (\ref{sol:steep}) coincides with the universal attractor discussed in Ref.~\cite{Kallosh:2013tua}. At the steep limit the solution converges on the universal attractor with no dependence on the gauge coupling, $\alpha$. In Ref. \cite{Channuie:2016iyy} the solution (\ref{sol:steep}) is also found for large curvature coupling cases, $\xi \gg 1$ in the NJL model. If we take the limit $\alpha\rightarrow +0$ in the gauged NJL model, the tensor-to-scalar ratio converges to the twice as large as the Starobinsky model. We conjecture the existence of new attractor in our model.

%%%
\subsection{Small gauge coupling limit}
%The CMB fluctuations are depend on the order of the limits, $\Lambda\rightarrow \infty$, and $\alpha\rightarrow 0$.

\begin{figure}[htbp]
  \centering
  \includegraphics[width=0.45\linewidth]{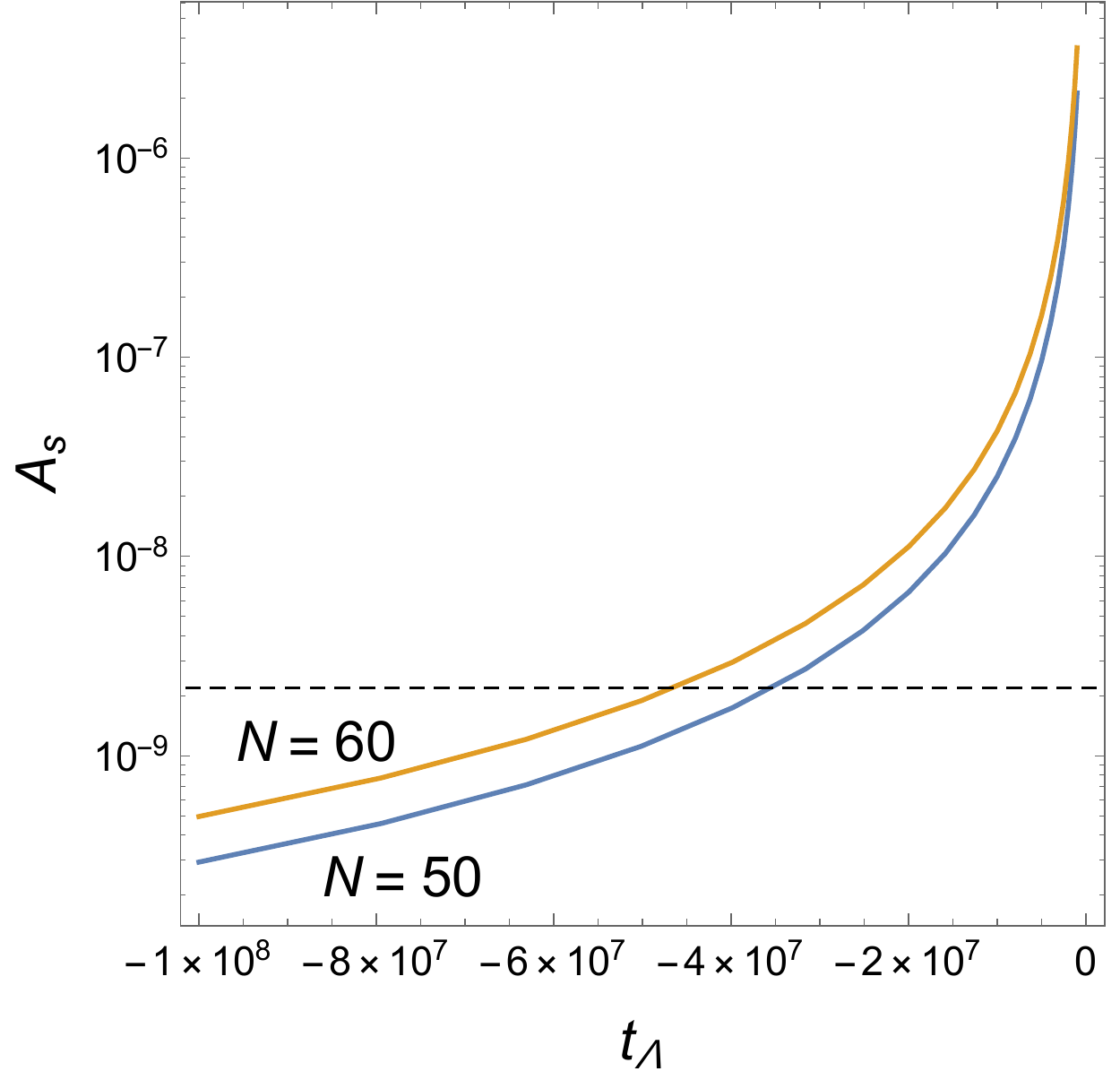}
  \caption{Curvature perturbation, $A_s$, as a function of $t_\Lambda=\ln(\mu/\Lambda)$ for $\alpha=0$, $G_{4r} = 10^{10}$, $a=4$ and $n=1$. The dashed line shows the observed value by Planck 2015.}
  \label{fig:As}
\end{figure}

The large compositeness scale limit of the CMB fluctuations obtained in this section do not reproduce the ones in Ref. \cite{Inagaki:2015eza}.
To see the difference we evaluate the effective potential at the $\alpha \to 0$ limit.
Taking the small coupling limit $\alpha \to 0$ of the effective potential $(\ref{einstein frame})$, it reduces to
% \begin{align}
%   &U = -\frac{\mu^2}{2t_\Lambda G_{4r}}\sigma^2 + \frac{\pi^2}{at_\Lambda^2}\left[3 + 2\ln\left(\frac{-at_\Lambda\mu^2}{8\pi^2\sigma^2}\right) - 4t_\Lambda\right]\sigma^4 + O(\alpha^2),   \label{epot:n1}\\
%   &\Omega^2 = 1 + \frac{1}{6}\left[1 - \frac{1}{12t_\Lambda}\left(1 + \ln\left(\frac{-at_\Lambda\mu^2}{8\pi^2\sigma^2}\right)\right)\right]\sigma^2 + O(\alpha^2),
%   \label{omega:n1}
% \end{align}
\begin{align}
  &U = -\frac{\mu^2}{2t_\Lambda G_{4r}}\sigma^2 + \frac{\pi^2}{at_\Lambda^2}\left[3 + 2\ln\left(\frac{-at_\Lambda\mu^2}{8\pi^2\sigma^2}\right) - 4t_\Lambda\right]\sigma^4 + O(\alpha),   \label{epot:n1}\\
  &\Omega^2 = 1 + \frac{1}{6}\left[1 - \frac{1}{2t_\Lambda}\left(1 + \ln\left(\frac{-at_\Lambda\mu^2}{8\pi^2\sigma^2}\right)\right)\right]\sigma^2 + O(\alpha),
  \label{omega:n1}
\end{align}
where we write $t_\Lambda=\ln(\mu/\Lambda)<0$. %The gauge coupling dependence appears form the first order, $O(\alpha)$.
In Fig. $\ref{fig:As}$ we plot typical behavior of the curvature perturbation, $A_s$, from the effective potential $(\ref{epot:n1})$ with $(\ref{omega:n1})$. We should fine tune the ration between the renormalization scale and the compositeness scale, $\mu/\Lambda \sim O(e^{-10^{7}})$, to acquire a observed value of the curvature perturbation, $A_s$. To obtain the observed value of the curvature perturbation without fine tuning we employ a larger gauge coupling, $\alpha$, here.

At the large compositeness scale limit, $\Lambda\rightarrow \infty$ the $\alpha$-independent terms in Eq.~$(\ref{epot:n1})$ disappear. The finite corrections appear from the next to leading terms,
\begin{align}
  &U = \frac{\alpha}{2\alpha_c} \left[ \frac{\mu^2}{G_{4r}}\sigma^2 + \frac{4\pi^2}{a}\sigma^4 \right] + O(\alpha^2)  \label{epot:largeL}\\
  &\Omega^2 = 1 + \frac{1}{6}\left[1 + \frac{\alpha}{2\alpha_c}\left(1 - \ln\left(\frac{a\alpha_c\mu^2}{8\pi^2\alpha\sigma^2}\right)\right)\right]\sigma^2 + O(\alpha^2).
\end{align}
Thus the $\alpha$-dependence of the effective potential has an decisive role for the CMB fluctuations. From this effective potential it is found that the observed value of the curvature perturbation, $A_s$, is generated for $\alpha\sim O(10^{-12})$ and $\mu\sim O(1)$. In Ref.~\cite{Inagaki:2015eza} the CMB fluctuations are evaluated only for the small gauge coupling $\alpha\sim O(10^{-12})$ and the larger gauge coupling case has not been discussed to avoid tuning of the renormalization scale, $\mu$.

% It should be noticed that the observed value, $A_s$, can be obtained by controlling not $\alpha$ but $\mu$. %It should be noticed that the observed value, $A_s$, can not be obtained for a larger $\alpha$.

%Fig.~\ref{fig:pert}, we show perturbation region and non-perturbation one.
% If we take the last term of \eqref{eq:zeta2} and \eqref{eq:lambda3} at $n\to 1$ limit, we have
% \begin{align}
%   equation
% \end{align}
% [When we discuss to calculate numerically in the next section, these terms are reproduced.]
% [If we consider a limit, $\Lambda\to\infty$, the another behaviour can be also taken, and this case coincides to the results in \cite{Inagaki:2015eza}. In this paper, this limit is prohibited by the finiteness of $\Lambda$, which distinguishes to ordinary NJL model.]

%----------------------------------------------------------------------------%

%----------------------------------------------------------------------------%
\section{Numerical results}
In this section we numerically calculate the CMB fluctuations, Eqs.~(\ref{epsilon:U}) - (\ref{N:U}). To confirm the arguments in the previous section we set the parameters ($\alpha$, $\Lambda$) to (0.5, 20) as a typical case.
%Unfortunately, we see that the results of analytic calculation in the perturbation region are excluded from observation, thus we also calculate including non-perturbation region.
%In Fig.~\ref{fig:pert}, we show perturbation region and non-perturbation one.
%\begin{figure}[htbp]
% \centering
%  \includegraphics[width=0.5\linewidth]{2n.pdf}
%  \caption{***}
%  \label{fig:pert}
%\end{figure}

\begin{figure}[htbp]
        \begin{minipage}{0.33\hsize}
                \centering
                \includegraphics[width=\linewidth]{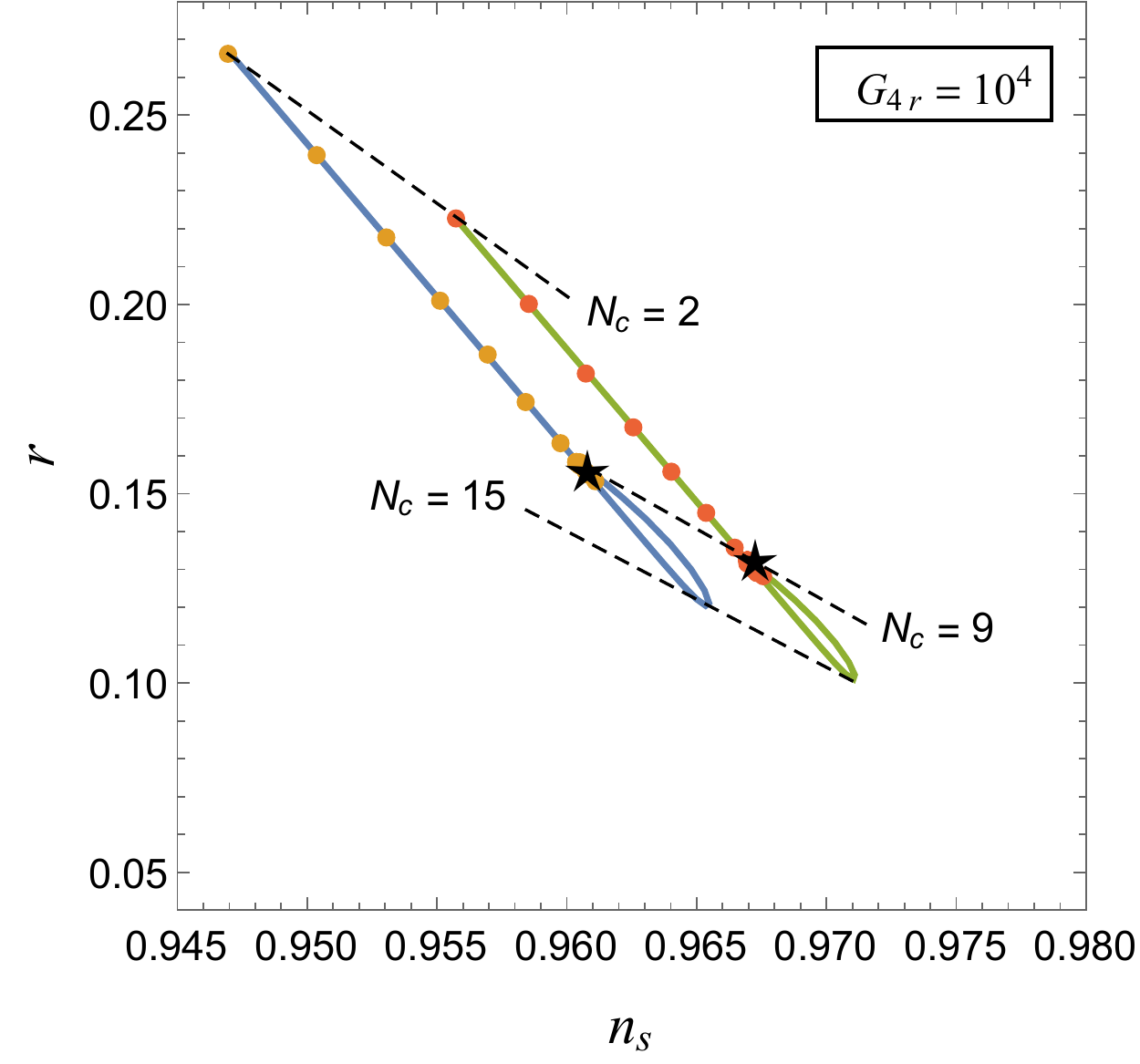}
        \end{minipage}
        \begin{minipage}{0.33\hsize}
        \centering
        \includegraphics[width=\linewidth]{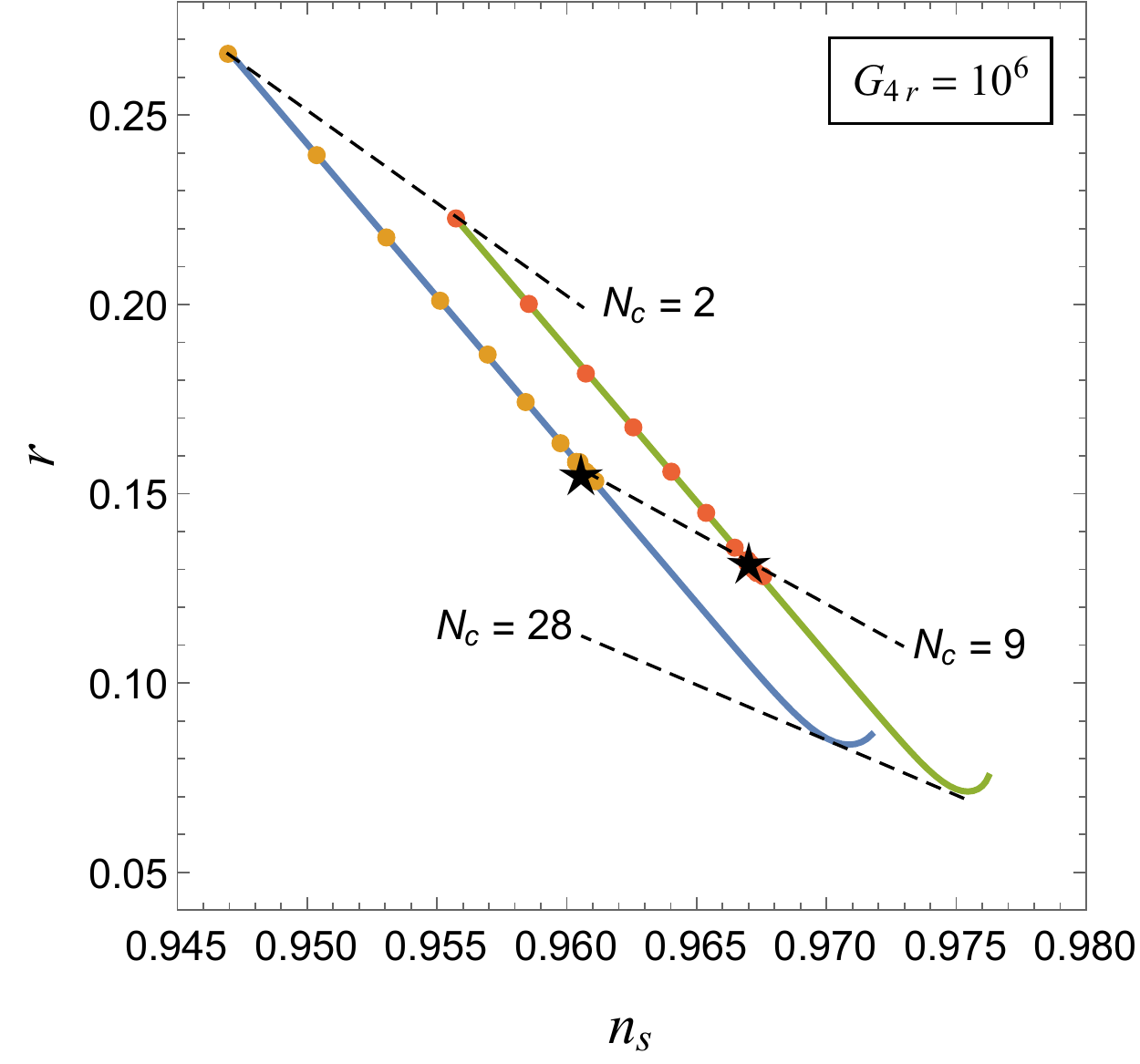}
        \end{minipage}
        \begin{minipage}{0.33\hsize}
                \centering
                \includegraphics[width=\linewidth]{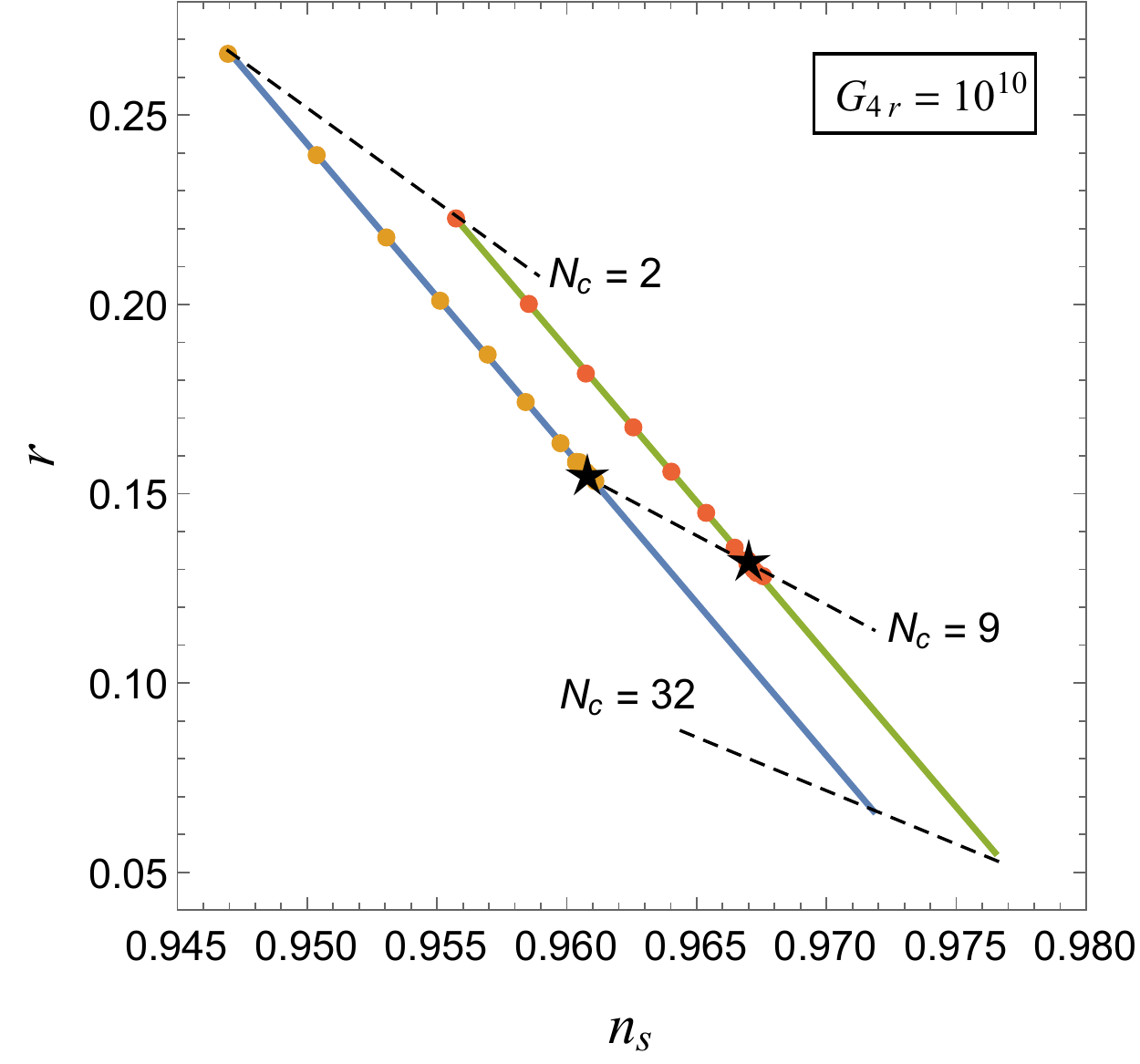}
        \end{minipage}
        \\
        \begin{minipage}{0.33\hsize}
                \centering
                \includegraphics[width=\linewidth]{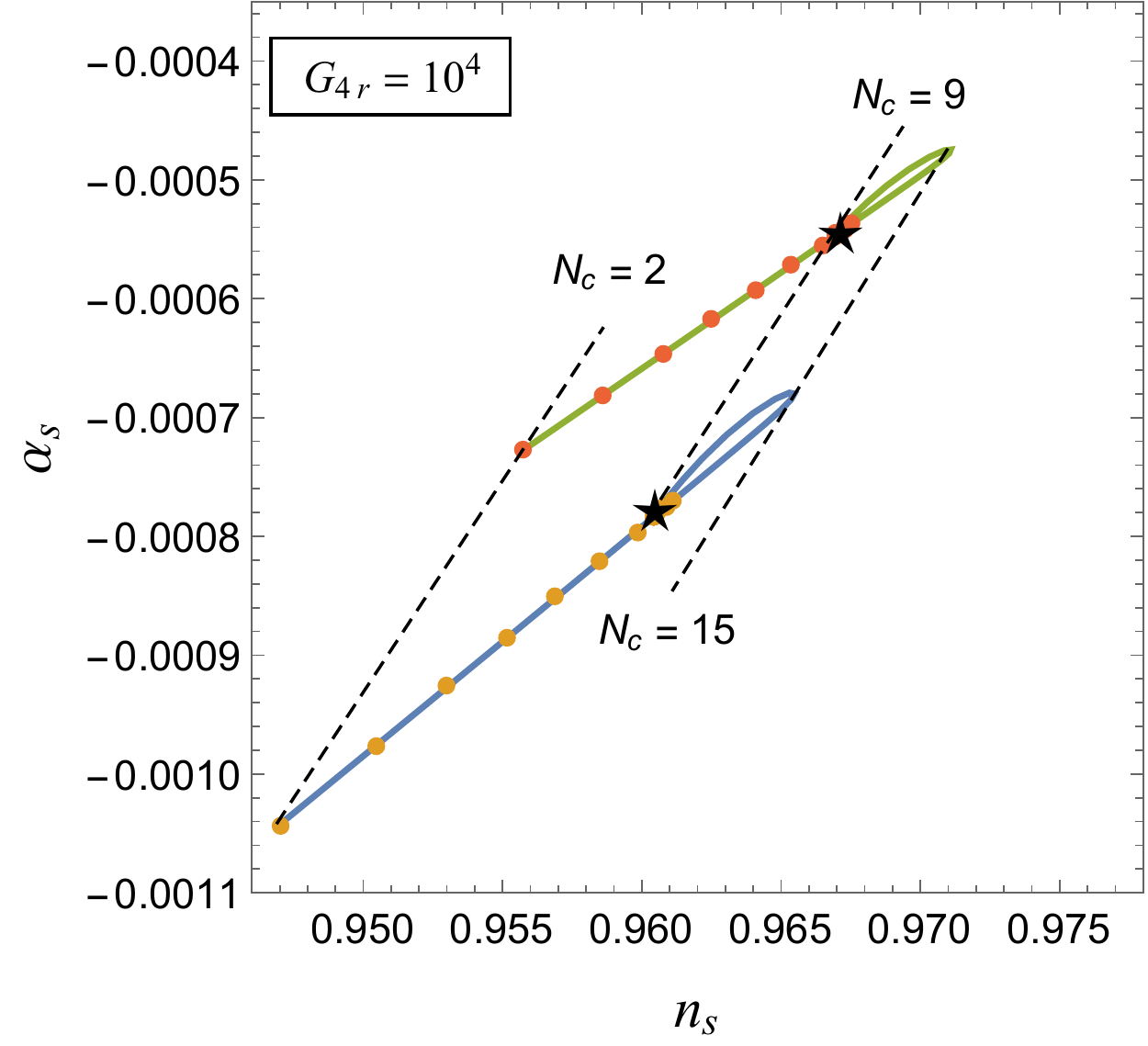}
        \end{minipage}
        \begin{minipage}{0.33\hsize}
        \centering
        \includegraphics[width=\linewidth]{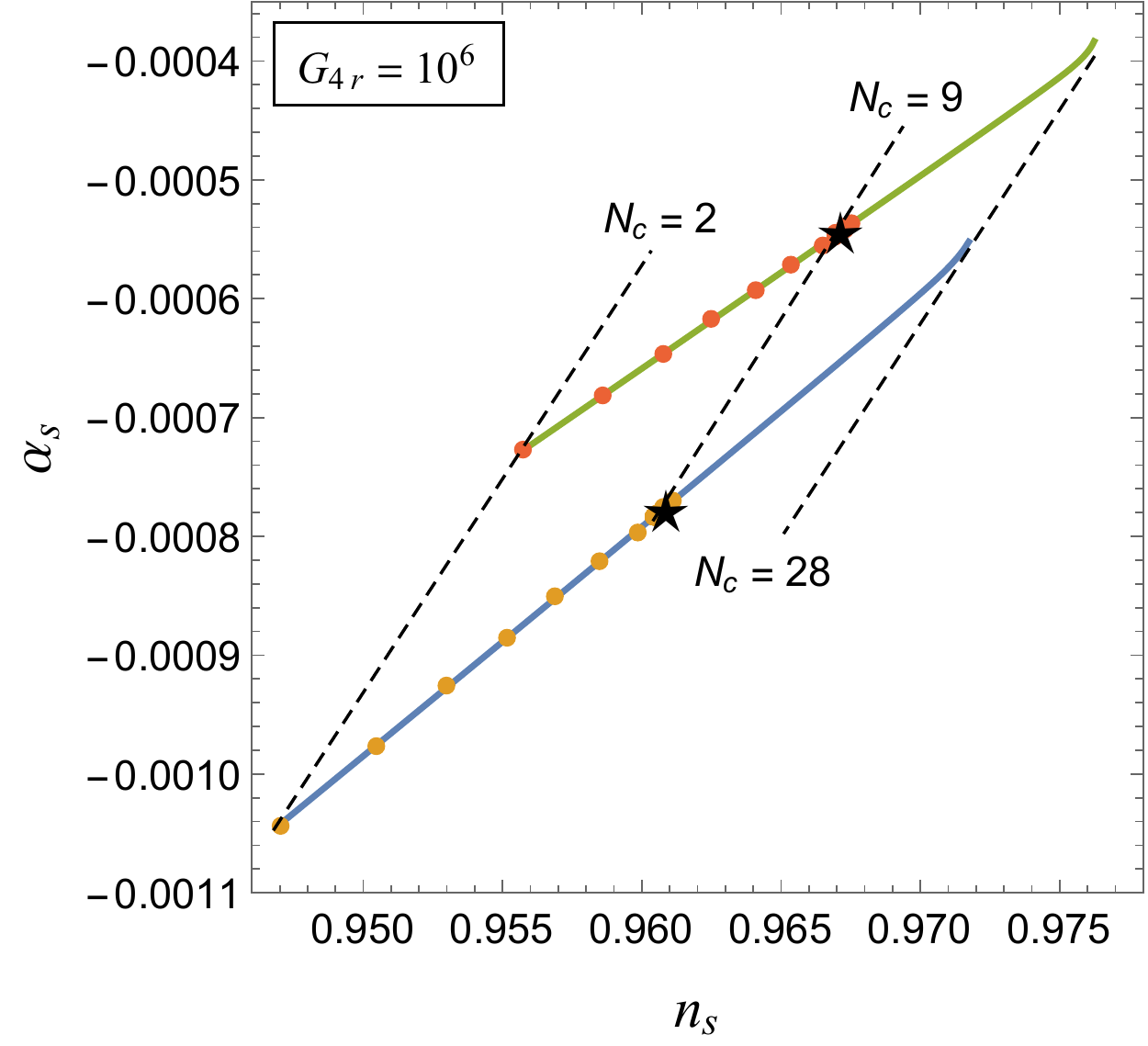}
        \end{minipage}
        \begin{minipage}{0.33\hsize}
                \centering
                \includegraphics[width=\linewidth]{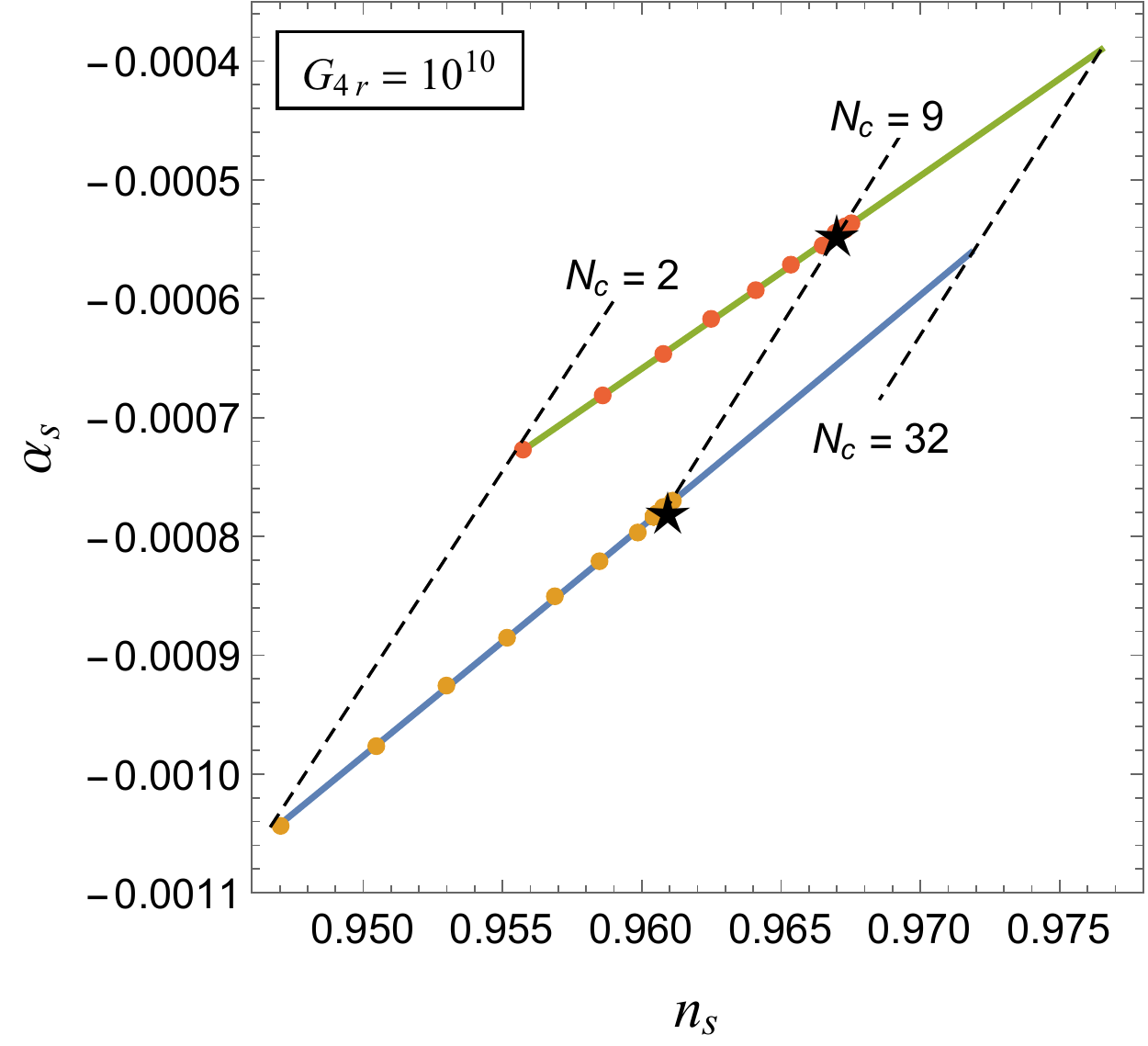}
        \end{minipage}
        \caption{Behavior of the tensor-to-scalar ratio, $r$, and the running of the spectral index, $\alpha_s$, as a function of the spectral index, $n_s$, for $N=50,60$ at $\alpha=0.5$, $\Lambda=20$ and $N_f=1$. The dashed lines is written to indicate the points with the equal $N_c$.}
        \label{fig:cmb fluctuation}
\end{figure}

In Fig. \ref{fig:cmb fluctuation} we plot the tensor-to-scalar ratio, $r$, and the running of the spectral index, $\alpha_s$, as a function of the spectral index, $n_s$, for $N_f =1$ with $N_c$ varies. The behavior in Fig. \ref{fig:cmb fluctuation} is consistent with the analysis in the previous section. The second term of $U$ in \eqref{einstein frame} has a dominant contribution between $N_c = 2$ and $N_c = 9$. In this interval the parameter $2n$ is grater than one, $2n > 1$, for $\alpha = 0.5$. The tensor-to-scalar ratio, $r$, and the running of the spectral index, $\alpha_s$, monotonically decreases and increases as a function of $n_s$, respectively. These results reproduce the one in the $\phi^{4n}$ chaotic inflation. For $N_c \gg 9$ the first term of $U$ in \eqref{einstein frame} gives a dominant contribution. The lines curve and the results approach to the ones for $N_c \sim 9$ $(2n=1)$ at the large $N_c$ limit. We indicate the point with $N_c \sim 9$ by the star.

\begin{table}[htb]
  \centering
  \begin{tabular}{l|ll}
     & $N=50$ & $N=60$ \\ \hline
    $2n = 1.70$ ($N_c = 2$) & $1.75\times 10^{-21}$ & $9.08\times 10^{-22}$ \\
    $2n = 1.52$ ($N_c = 3$) & $3.37\times 10^{-13}$ & $2.11\times 10^{-13}$ \\
    $2n = 1.38$ ($N_c = 5$) & $1.65\times 10^{-8}$ & $2.67\times 10^{-10}$ \\
    $2n = 1.03$ ($N_c = 8$) & $3.12\times 10^{-6}$ & $2.58\times 10^{-6}$ \\
    $2n = 0.50$ ($N_c = 25$) & $6.74\times 10^{-5}$ & $6.15\times 10^{-5}$ \\
  \end{tabular}
  \caption{The renormalization scale $\mu$ corresponding to the observable of curvature perturbation for $N_c = 2, 3, 5, 8, 25$ and the e-folding number $N = 50, 60$ at $\alpha = 0.5$, $N_f=1$ and $G_{4r} = 10^{10}$. The mass scale is normalized by $M_p=1$. }
  \label{tab:mu}
\end{table}

In these calculations we fix the renormalization scale, $\mu$, to generate the observed curvature perturbation, $A_s$. We show the scale in Table \ref{tab:mu} at $\alpha = 0.5$, $\Lambda=20$ and $G_{4r} = 10^{10}$ for $N_c=2, 3, 5, 8$ and $25$.
Similar behavior is obtained with the one at the flat limit. The parameter, $n$, is defined as a function of $N_c$ and $\alpha$ in \eqref{def:n}. The $\alpha$-dependence of the renormalization scale is also observed in the Table \ref{tab:mu}. The parameter, $n$, is monotonically decreasing as a function of $\alpha$. Thus, the scale $\mu$ decreases as the gauge coupling, $\alpha$, increases.

\begin{figure}[htbp]
\begin{center}
        \begin{minipage}{0.44\hsize}
                \centering
                \includegraphics[width=\linewidth]{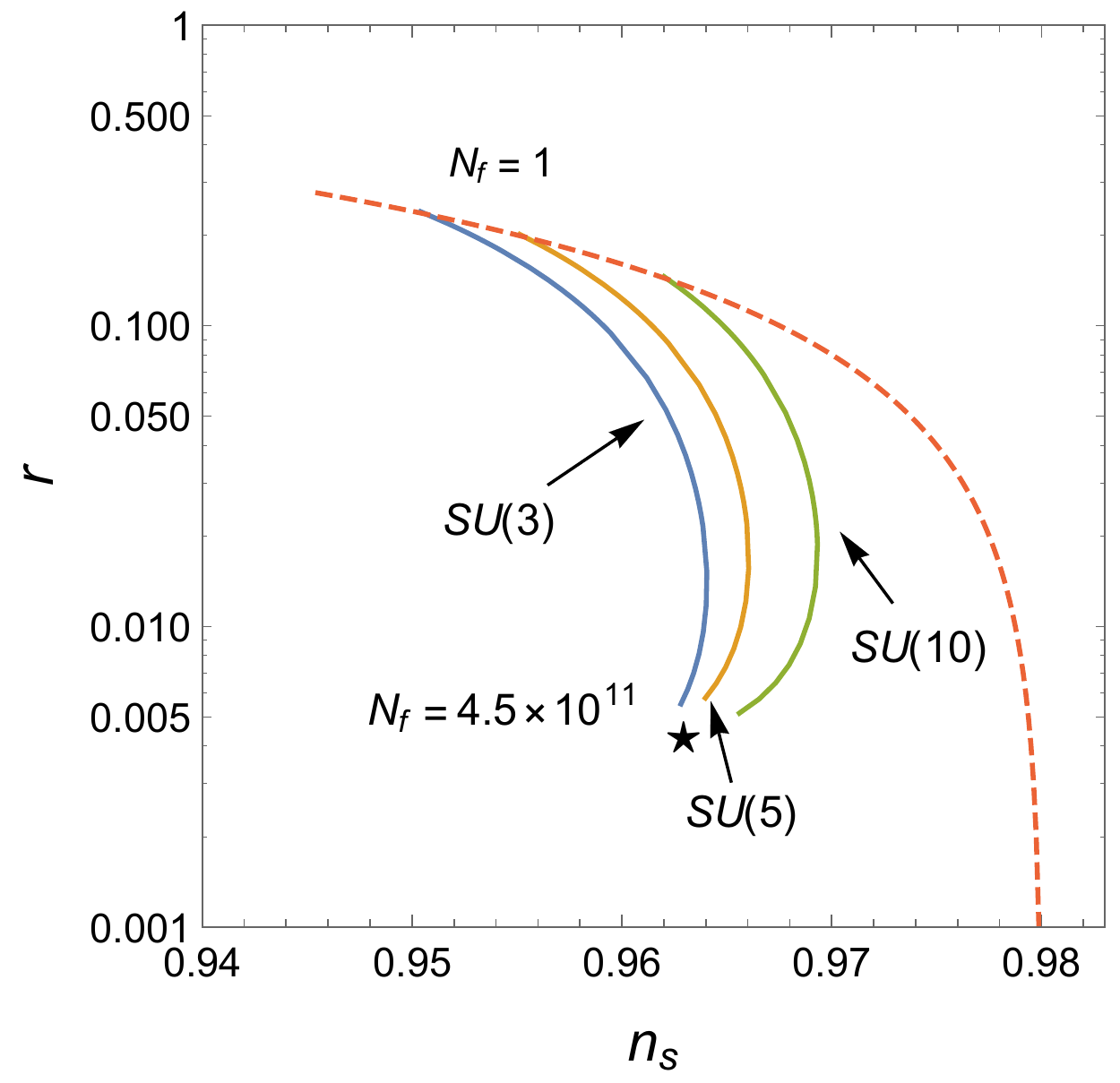}
        \end{minipage}
        \hspace{4mm}
        \begin{minipage}{0.44\hsize}
        \centering
        \includegraphics[width=\linewidth]{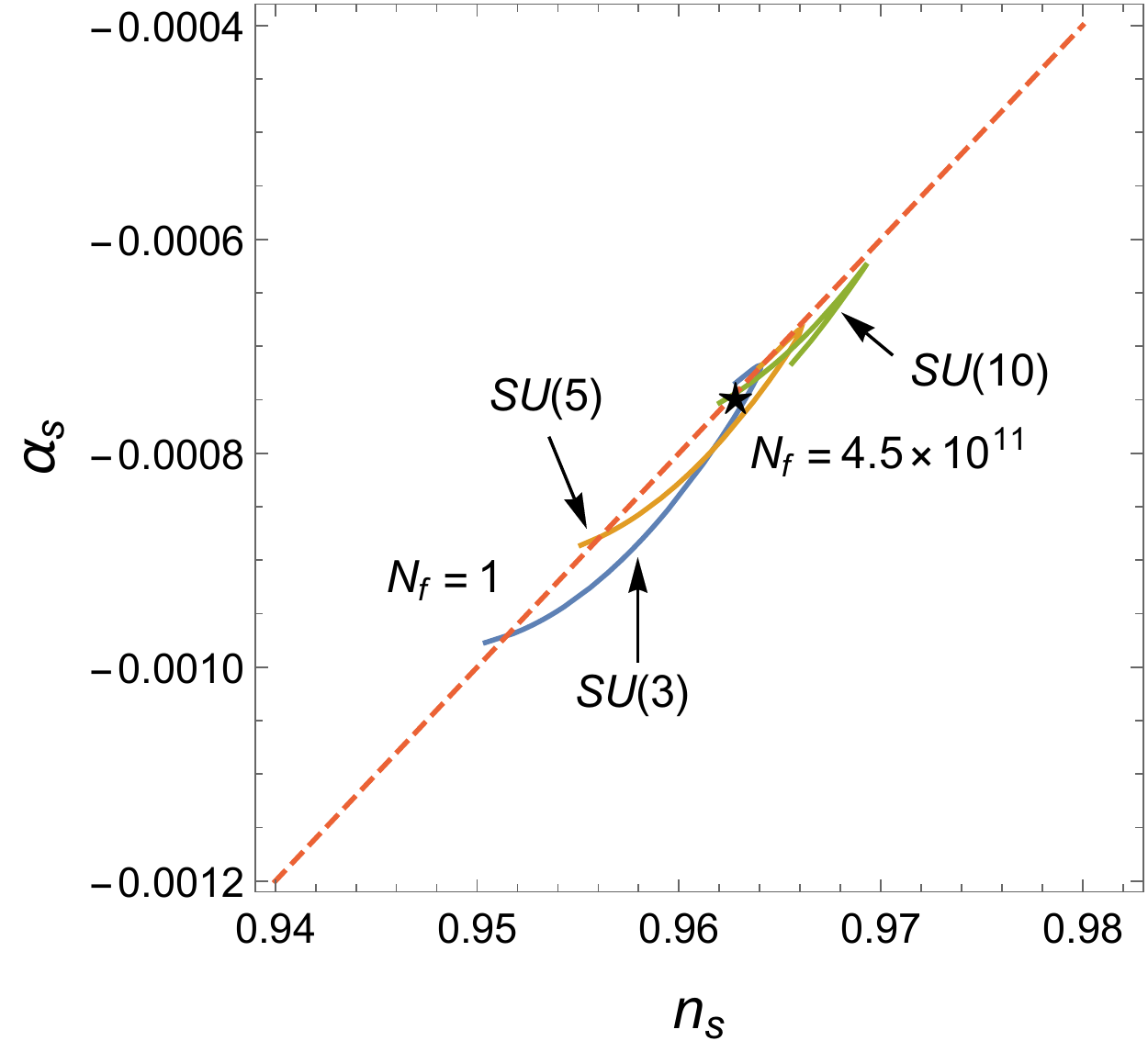}
        \end{minipage}
        \caption{Behavior of the tensor-to-scalar ratio, $r$, and the running of the spectral index, $\alpha_s$, as a function of the spectral index, $n_s$, for $N=50$ at $\alpha=0.5$, $\Lambda=20$. The dashed lines is written to indicate the flat limit. The star indicates the steep limit.}
        \label{fig2:cmb fluctuation}
        \end{center}
\end{figure}

Next we evaluate the $N_f$ dependence on the tensor-to-scalar ratio, $r$, and the running of the spectral index, $\alpha_s$. In Fig.~\ref{fig2:cmb fluctuation} we draw $r$ and $\alpha_s$ as a function of $n_s$ for $N_c=3, 5$ and $10$ with $N_f$ varies. The behavior depends on the gauge group $SU(N_c)$. A larger spectral index, $n_s$, is obtained as $N_c$ increases. For $N_f=1$ the results can be well described by Eq.~(\ref{limit:flat}) at the flat limit. It is observed that the results converge to the attractor point (\ref{sol:steep}) at the large $N_f$ limit. The results are consistent with the observation by Planck 2015 for a large number of fermion flavors.

In Table \ref{tab:Nf-Lambda} we show the CMB fluctuations for $N_c = 2$, $N=50$, $\alpha = 0.5$ and $G_{4r} = 10^{10}$ as $\Lambda$ and $N_f$ vary. All the cases are indistinguishable in the accuracy of calculations under discussion. The CMB fluctuations have no large dependence on the compositeness scale, $\Lambda$, and the number of the fermion species, $N_f$.
%%%
In the original NJL model the composite scale corresponds to the QCD scale. It is at most $10-100$ times larger than the renormalization scale, i.e. meson scale. Here we regard the model as a scale up of QCD like model. However, the different dynamics may induce a larger compositeness scale. Thus we consider the large compositeness scale. Though the scale, $\Lambda\sim 2000$, is too large, it is also useful to verify that the results do not converge to the ones obtained at the large compositeness scale limit in Ref.~\cite{Inagaki:2015eza}.

The inflationary parameters values approach to the universal attractor for an extremely large $N_f$. Although the case, $N_f = 4.5\times 10^{11}$, is unnatural and not consistent with the constant $SU(N_c)$ gauge coupling assumption, we include the case to observe the universal attractor at the steep limit.
\begin{table}[htb]
  \centering
  \begin{tabular}{l|l|lll}
    $N_f$ & $\Lambda$ & $n_s$ & $r$ & $\alpha_s$ \\ \hline
    $N_f = 1$ & $\Lambda = 20$ & $0.947$ & $0.27$ & $-0.0010$ \\
    $N_f = 1$ & $\Lambda = 200$ & $0.947$ & $0.27$ & $-0.0010$ \\
    $N_f = 1$ & $\Lambda = 2000$ & $0.947$ & $0.27$ & $-0.0010$ \\
    $N_f = 11$ & $\Lambda = 20$ & $0.947$ & $0.27$ & $-0.0010$ \\
    $N_f = 21$ & $\Lambda = 20$ & $0.947$ & $0.27$ & $-0.0010$\\
    $N_f = 31$ & $\Lambda = 20$ & $0.947$ & $0.27$ & $-0.0010$\\
    \hline
    $N_f = 4.5\times 10^{11}$ & $\Lambda = 20$ & $0.964$ & $0.0048$ & $-0.00076$\\
  \end{tabular}
  \caption{CMB fluctuations for  $N_c = 2$, $N=50$, $\alpha = 0.5$ and $G_{4r} = 10^{10}$.}
  \label{tab:Nf-Lambda}
\end{table}

%----------------------------------------------------------------------------%
\section{Reheating temperature}
After the inflationary expansion cooled down the Universe, the potential energy of the inflaton is released through the decay process of the inflaton into light particles. The reheating temperature depends on the decay modes. Here we assume a Yukawa-type interaction $y_h \bar{\Psi}\Psi\sigma$ between the composite scalar $\sigma$ and a light fermion field, $\Psi$ with the coupling constant $y_h$. In a standard scenario of the chaotic inflation the reheating temperature is roughly estimated by \cite{Kofman:1997yn}
\begin{align}
  T_r \sim 0.2 \sqrt{\frac{y_h^2}{8\pi}M M_{p}},
\end{align}
where $M$ represents the effective $\sigma$ mass at the end of the inflation. The composite field $\sigma$ oscillates around the origin of the effective potential after the inflation era. The typical scale of the effective mass is given by
\begin{align}
  M^2 = \left.\frac{\partial^2 V_E}{\partial \sigma^2}\right|_{\sigma = \sigma_*} ,
\end{align}
where $\sigma_*$ is the amplitude of oscillations.

In Table \ref{tab:tr} we set $\sigma_*=1 (< \sigma_{end}=O(1))$ and evaluate the reheating temperature. Since we evaluate the decay process of the inflaton at tree level with respect to the coupling $y_h$, the results is valid for $y_h <1$. In the perturbative regime the upper limit of the reheating temperature is below the order $10^{15}$GeV for $\alpha = 0.5$ and $G_{4r} = 10^{10}$.
\begin{table}[htb]
  \centering
  \begin{tabular}{l|ll}
     & $N=50$ & $N=60$ \\ \hline
    $2n = 1.70$ ($N_c = 2$) & $4.9 y_h\times 10^{13}$GeV & $4.4 y_h\times 10^{13}$GeV\\
    $2n = 1.52$ ($N_c = 3$) & $5.9 y_h\times 10^{13}$GeV & $5.3 y_h\times 10^{13}$GeV\\
    $2n = 1.38$ ($N_c = 5$) & $2.1 y_h\times 10^{14}$GeV & $6.0 y_h\times 10^{13}$GeV\\
    $2n = 1.03$ ($N_c = 8$) & $9.4 y_h\times 10^{13}$GeV & $8.6 y_h\times 10^{13}$GeV\\
    $2n = 0.50$ ($N_c = 25$) & $1.8 y_h\times 10^{12}$GeV & $1.8 y_h\times 10^{12}$GeV\\
  \end{tabular}
  \caption{Reheating temperature for $N_c = 2, 3, 5, 8, 25$ and the e-folding number $N = 50, 60$ at $\alpha = 0.5$ and $G_{4r} = 10^{10}$.}
  \label{tab:tr}
\end{table}

%----------------------------------------------------------------------------%
\section{Conclusion}
In this paper we investigated the gauged NJL model with a finite compositeness scale as an alternative scenario of Higgs inflation.
We assume that the potential energy of a composite scalar field causes the exponential expansion of the Universe at inflation era. Hence the CMB fluctuations are produced from the quantum fluctuation of the composite scalar.

There is a solution of the RG equations where the gauged NJL model coincides with the gauge-Higgs-Yukawa theory at the compositeness scale. The effective potential is calculated by solving the renormalization group equations below the compositeness scale. In our model the potential depends on the gauge coupling, the four-fermion coupling, the number of the gauge group, the number of the fermion species, the renormalization scale and the compositeness scale. We evaluated the CMB fluctuations starting from the obtained potential.

Analytic results are obtained at the specific limits. Taking the flat limit, the gauged NJL model behaves as the $\phi^{4n}$ chaotic inflation. The exponent $n \in (0, 1)$ is given by the gauge coupling, $\alpha$, and the rank of the gauge group, $SU(N_c)$. It is found that the tensor-to-scalar ratio generated in the model with a large $n$ $>1/2$ is larger than the upper limit by Planck 2015.  At the steep limit the effective potential of the model coincides with the Starobinsky inflation which is equivalent to Higgs inflation by Bezrukov and Shaposhnikov. Hence, the solution reaches the universal attractor with no dependence on the gauge coupling. It is quite remarkable that finite scale gauged NJL model inflation interpolates between above two well-known inflationary universe models.

We numerically evaluated the CMB fluctuations as the model parameters vary and found large dependence on the gauge group $SU(N_c)$ and the number of the fermion flavors $N_f$. A smaller tensor-to-scalar ratio is generated at larger $N_f$.
Therefore we found a parameter range of the gauged NJL model to generate CMB fluctuations consistent with the observation by Planck 2015. In the possible parameter range the product $\alpha N_c$ is larger than unity. On the other hand we employ the one-loop effective potential which is perturbatively calculated with the expansion parameter, $\alpha N_c$.
%%%
Fermion loop corrections in the gauge boson vacuum polarization proportional to $\alpha N_c$. It contributes the screening of the gauge interaction. Since we drop the running effect of the gauge coupling here, we expect that the fermion loop corrections have no large contribution to the CMB fluctuations even for $\alpha N_c >1$. It is one of our remaining problems to include the running of the gauge coupling with the fermion loop corrections. It may modify the gauge coupling dependence of the results. However, we expect that the results show some characteristic features of the gauged NJL inflation. To study the further detail of NJL inflation, we need to develop a procedure beyond the perturbation for $\alpha N_c$. This will be discussed elsewhere.

The important remark is in order. Even we consider the compositeness scale of the model, though finite, to be  much larger than the Planck scale, we cannot avoid to comment on the quantum gravity effects beyond the Planck scale. However, this depends very much from the quantum gravity theory under discussion. For instance, if we consider stringy quantum gravity, one should extend the number of gravitational partners. From other side, it  is pointed out that the scalar type quantum fluctuations dominate much more than the tensor type ones in a two-loop conformal gravity \cite{Hamada:2015ywk}. Although it is the result in a specific renormalizable model of the quantum gravity, we expect that the purely quantum gravity effects do not enhance the tensor-to-scalar ratio. In higher-derivative multiplicative quantum gravity the effect will be in the inducing of $R^2$ term in the action. However, its quantum-induced coefficient will be very small so it plays no essential role in our considerations. In the gauged NJL model under consideration the quantum gravity effects also may slightly modify the RG flow and then  the corresponding effective potential. However, we consider the particle physics model with a large number of fermion species. At the large $N_f N_c$ limit the perturbative quantum gravity effects can be neglected compared with the fermion loop corrections at the equal order of the loop expansion. The corresponding effects are higher order corrections and perturbatively suppressed in a sense of the loop expansion. This is similar to conformal anomaly which is proportional to number of fields (i.e. the fermion fields number, $N_fN_c$, proportional sector plus the gauge fields number proportional sector). It introduces $R^2$ terms in the effective action \cite{Buchbinder:1992rb} and contributes to approach the fluctuation to the steep limit for large $N_fN_c$. Note that this is customary in the inflationary cosmology induced by matter quantum effects which are always considered to be dominant compare with quantum gravity effects even if the inflation scale is much higher than Planck scale.
%----------------------------------------------------------------------------%

\section*{Acknowledgements}
The work by TI is supported in part by JSPS KAKENHI Grant Number 26400250 and that by SDO is supported in part by MINECO (Spain), project FIS2013-44881 and by MES (Russia).
%----------------------------------------------------------------------------%
%----------------------------------------------------------------------------%

\end{document}